\PassOptionsToPackage{bookmarks=false}{hyperref}
\documentclass[letterpaper, 10 pt, conference]{IEEEtran}  
                          
\IEEEoverridecommandlockouts                              

\makeatletter
\newcommand*{\rom}[1]{\expandafter\@slowromancap\romannumeral #1@}
\makeatother
\usepackage{transparent}
\usepackage{amsmath}
\usepackage{algorithm}
\usepackage{algpseudocode}
\usepackage{hyperref}
\usepackage{graphicx}
\usepackage{caption}
\usepackage{textcomp}

\usepackage{afterpage}

\hypersetup{
    colorlinks=true,
}

\def\BibTeX{{\rm B\kern-.05em{\sc i\kern-.025em b}\kern-.08em
    T\kern-.1667em\lower.7ex\hbox{E}\kern-.125emX}}
\title{\LARGE \bf
Decentralized Patient Centric e-Health Record Management System using Blockchain and IPFS
}

\author{\IEEEauthorblockN{ Gaganjeet Singh Reen} \\
\IEEEauthorblockA{\small Network Security \& Cryptography Lab \\
\small IIIT Allahabad\\
\small Allahabad, India \\
reen.gagan@gmail.com}
\and
\IEEEauthorblockN{ Manasi Mohandas} \\
\IEEEauthorblockA{\small Network Security \& Cryptography Lab \\
\small IIIT Allahabad\\
\small Allahabad, India \\
manasi.mds@gmail.com}
\and
\IEEEauthorblockN{ S. Venkatesan} \\
\IEEEauthorblockA{\small Network Security \& Cryptography Lab \\
\small IIIT Allahabad\\
\small Allahabad, India \\
venkat@iiita.ac.in}
}

\begin{document}
\onecolumn
\noindent \textcopyright 2019 IEEE. Personal use of this material is permitted. Permission from IEEE must be obtained for all
other uses, in any current or future media, including reprinting/republishing this material for advertising
or promotional purposes, creating new collective works, for resale or redistribution to servers or lists, or
reuse of any copyrighted component of this work in other works. \newline

\noindent Accepted to be published in 2019 IEEE Conference on Information and Communication Technology.

\noindent DOI: 10.1109/CICT48419.2019.9066212
\thispagestyle{empty}
\newcommand\blankpage{
    \null
    \thispagestyle{empty}
    \addtocounter{page}{-1}
    \newpage
    }
\twocolumn

\maketitle

\thispagestyle{empty}
\pagestyle{empty}

\begin{abstract}

Electronic Health Records(EHR) are gaining a lot of popularity all over the world.
The current EHR systems however have their fair share of problems related to privacy and security. We have proposed a mechanism which provides a solution to most of these problems. Using a permissioned Ethereum blockchain allows the hospitals and patients across the world to be connected to each other. Our mechanism uses a combination of symmetric and asymmetric key cryptography to ensure the
secure storage and selective access of records. It gives patients full control over their health records and also allows them to grant or revoke a hospital\textquotesingle s access to his/her records. We have used IPFS(inter planetary file system) to store records which has the advantage of being distributed and ensures immutability of records. The proposed model also maintains the statistics of  diseases without violating the privacy of any patient.\smallskip


\end{abstract}

\section{INTRODUCTION}

In 2014, 76\% of hospitals in the United States had adopted basic EHR systems [5]. While a lot of hospitals in the developing nations still lack the facilities required to shift to an EHR system, the adoption of EHR has been on a rise globally. The traditional EHR systems however have a lot of problems when it comes to privacy, security and ease of use. The records are completely controlled by the hospitals that created the record. These records can be misused if the hospital does not have adequate security measures in place. There is also a considerable lack of uniformity in the records that are created and stored. The records stored by one hospital might not be accessible or may be incomprehensible to the doctors at another hospital. This primarily occurs because different hospitals use different providers to design their EHR software systems. 

Recently there have been several  ransomware attacks in which the hackers install malware on the medical organization’s servers. They make the data inaccessible and release it only once their demands have been met. 
Two recent incidents \cite{c6} at Howard University Hospital, Washington proved that adequate data security measures are required in the field of e-Health Records. On May 14, 2013, one of the hospital’s medical technicians was charged with violating HIPAA(Health Insurance Portability and Accountability Act), as she had sold private patient information. In another incident in the same hospital, a contractor had downloaded nearly 34000 patient’s files to his personal laptop.

In India, National eHealth Authority(NeHA) \cite{c7} is a proposed organization which aims to set certain regulations in the field of eHealth care. This is being done to ensure standardization of eHealth records which would facilitate
secure access to these records while ensuring privacy. The model proposed in this paper could be used for this very purpose.
In order to provide absolute privacy and security, we have put to use the concepts of cryptography, blockchain and IPFS.
Blockchain is an immutable list of records, which are stored in blocks with each block being linked to the previous block. In a Blockchain every transaction made between the accounts including between a user and a smart contract account is stored in a block. These transactions are public and irreversible once they have been stored in a block. So everyone can see who has accessed which smart contract.

We have used the Ethereum blockchain because it enables us to use smart contracts. We are making use of a permissioned version where select participant nodes are miners. The actual record data is stored off the blockchain and on the network nodes which are a part of IPFS(Inter Planetary File System).\smallskip

\section{COMPARISON WITH EXISTING WORKS} 

MedRec \cite{c1} is one of the earliest suggested models for a blockchain based medical record management system. It utilizes the Ethereum Blockchain and Smart Contracts to store accessibility details of the health record. The actual health record is not stored on the Blockchain, it is stored on the Healthcare Providers database which is operated by a third party. Hence these records are still susceptible to attack or misuse. The primary difference between MedRec and our approach is that we store the records in a distributed manner and do not rely on a third party. Also, the use of hashes in IPFS guarantees the immutability of records.\smallskip

 In Blockchain for Healthcare \cite{c2}, the patient data is stored on the blockchain after being encrypted with the patients private key. This data can be decrypted using the patients public key which is provided to users such as hospitals and researchers to whom the patient has given permission. This is in stark contrast to our approach in which the patients have complete control over who can view their data. In our model, the patients private key is required to decrypt the data instead of encrypting it. Also, we do not store the data on the blockchain but instead use IPFS for the same.
 
 Blockchain For Health Data and Its Potential Use in Health IT and Health Care Related Research \cite{c3} makes use of the blockchain to store encrypted medical record details such as where it is stored(hashed pointers) and who has access to it. Only the owner of the data can change the access control policies. All the data is stored in a data lake. For identity authentication the suggested method in this model is a biometric system which would be more secure than a password.The difference between this approach and ours is the distributed nature in which the data is stored in our system.
 
Qi Xia et al \cite{c20} proposed the blockchain based health record sharing model using the cryptographic keys and smart contracts. Alevtina Dubovitskaya et al. \cite{c21}, has proposed a secure and trustable health record sharing method based on Blockchain technology. They concentrated on the Oncology department of hospitals and the assumption is access to all records or no access at all. 
It is not possible for a patient to disclose some of the records because the same key is used to encrypt all records.
In our approach, the patient can specify which records it wants the hospital to gain access to.

Considering the existing works and their shortcomings, this paper proposes a secure and efficient model to store and access the health record in the presence of the outside attackers.


 
 

\section{REQUIREMENTS AND PRELIMINARY}
The main requirements that need to be satisfied in an electronic medical record management system are-
\begin{itemize}
\item The patient must have complete control over who can view his/her records.
\item The patient should be able to grant and revoke access to his/her records as and when he/she deems fit.
\item The patient records once stored should be immutable and changing them should be infeasible.
\item The hospital should not be able to deny creating the record at a later point of time and the patient should not be able to claim that a record was created in his/her absence.
\end{itemize}
In this EHRS model there are primarily two types of users: Patient and Hospital. The patient is the only type of user who can grant and revoke access to the health records.


The hospital enters the health record details on the website where it is encrypted and stored in IPFS using the patients public key (\textit{P\textsubscript{Pub}}). Hence, it can be decrypted by using the patients private key (\textit{P\textsubscript{Priv}}) only. If the hospital wishes to have access to this record they can request the patient to give them access. This access can be revoked at any point of time. The hospital has a database which stores the patient name patient ID and contact details.

\section{PROPOSED MODEL}
All the participating nodes on the network i.e. hospitals, patients, Insurance agency, etc. are required to be a part of two networks- the permissioned blockchain and IPFS. However not all of them are not required to keep the entire blockchain stored(light nodes).

\begin{figure} [!htb]
	\centering
	\fbox{
		\parbox{24em}{
			Health Record Structure: \\
			\textit{
				\phantom{Vot}Patient Id \\
				\phantom{Vot}Gender \\
				\phantom{Vot}Age \\
				\phantom{Vot}Disease\\
				\phantom{Vot}Diagnosis \\
				\phantom{Vot}Location \\
				\phantom{Vot}Medication \\
				\phantom{Vot}Suggestion \\
				\phantom{Vot}Next Review \\
				\phantom{Vot}Notes \\
				\phantom{Vot}Date \\
				\phantom{Vot}Doctor's Name \\
				\phantom{Vot}Hospital Id \\
			}
		}
	} 
	\caption{\label{fig1:The-caption}Structure of Health Record}
\end{figure}

In a Permissioned Blockchain a group of participants are decided in advance who in turn validate blocks and transactions. The miners do not need any incentive to mine blocks in this model of the blockchain. Thus the concept of gas cost has no significance in a permissioned blockchain. In our model, the hospitals will be the miners. The hospitals should have no problem agreeing to spend heir computational power and mine blocks because in the absence of such a system, the hospitals would have had to pay large sums of money to the organisations handling their data and interfaces. If the concept of gas and ether does not play a role in the blockchain, it is apparent that Proof of Stake or Proof of Work cannot be used as consensus protocols. As a result, for our use case, we propose using \textbf{Proof of Authority}\cite{c15} as a suitable consensus protocol.
A Permissioned Blockchain serves as an immutable ledger for all the activities that take place on the network. The term activities used here refers to actions like :-
\begin{itemize}
    \item A hospital creating a record for a patient.
    \item A patient accessing a record.
    \item A patient granting privileges to or revoking privileges of a hospital to view his/her records.
    \item A hospital accessing a record.
\end{itemize}
The sample structure of a health record is shown in Fig. 1, and Fig 2. is a representation of the proposed system.
There are three smart contracts in our proposed system :-

\begin{itemize}
	\item The first one contains a mapping from \textit{P\textsubscript{Pub}} to the hashes of the patient's records. 
	\item The second one contains a mapping from the combined patient and hospital public key (\textit{PH\textsubscript{Pub}}) to the record that the hospital has access to.
	\item The third smart contract is used to store the number of cases of each disease being reported.
\end{itemize}

\begin{figure*} [!htb]
\centering
\includegraphics[width=1\textwidth,height = 10cm]{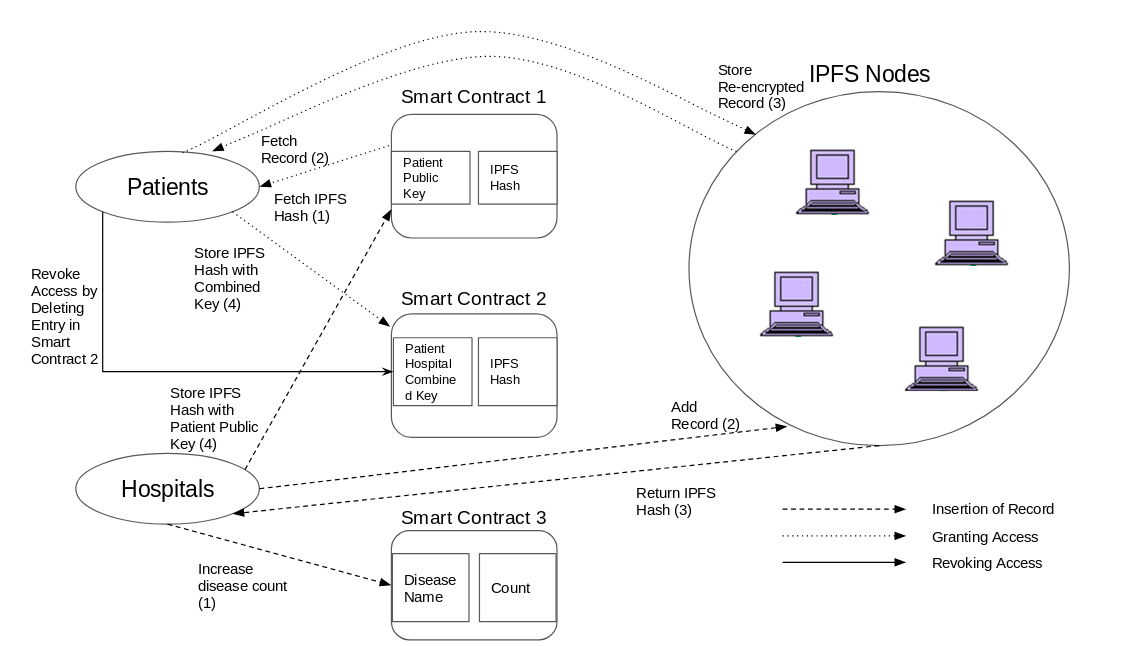}
\centering \caption{\label{fig1:The-caption}Architecture of e-Health Record Management System }
\end{figure*}

Any hospital wishing to access patient records will have to do so by sending a request to the patient and the patient in turn will grant access by encrypting the records with the hospital public key (\textit{H\textsubscript{Pub}}) and storing the hash of the record in the second smart contract mentioned above. This storage will be registered as a transaction to the second smart contract. As a result, everyone will know which hospital has access to the records of which patient.
Also, when a hospital creates a record, a corresponding transaction is created to the first smart contract which is again visible to all the participating nodes on the system.

In IPFS data can be accessed only using the SHA-256 hash value stored in base58 format. So without the hash, data is inaccessible.
The hashing operation can be represented as \newline \textit{f\{0,1\}\textsuperscript{*} $\to$ \{0,1\}\textsuperscript{n}} where f is the hash function and n = 256(because the hashing algorithm used is SHA 256).\newline
In IPFS each data entry is stored in the form of a merkle DAG so any changes in the data would not be possible as the hash value will change. Hence it is tamper resistant which is a desired property for health records.

\subsection{SignUp Phase}
\subsubsection{User SignUp}
The SignUp phase is required so that users can access their records and grant and revoke access from machines other than the ones they used to become a part of the blockchain. Signing up allows them to access their private keys from systems other than their own. The user must however make sure to clear the browser cache and history after accessing the system from a machine which can't be trusted. \newline
We also need to ensure that a particular patient joins the blockchain just once i.e. a single patient should not be associated with multiple accounts. To ensure this, the private keys of the patients should be derived from something that is unique to a patient like a fingerprint\cite{c17}.\newline
In order to signup and become a part of the network, the user must join the permissioned Ethereum blockchain by providing his/her fingerprint so as to derive the Ethereum blockchain private key\footnote{This private key is not directly stored anywhere}. The public key of the node(which is used to encrypt the records) is extracted from this particular private key. The combination of the public key and the private key is encrypted with the patients fingerprint(biometric encryption) and is then stored securely in a database.



We use both symmetric key cryptography as well as asymmetric key cryptography.The symmetric key cryptography is
where the decryption password (derived from the fingerprint of the user) comes into play. The user provides his fingerprint and the password derived from it is used to encrypt a file containing the public key of the user stored in conjunction with the private key of the user. This encrypted file is stored in the local storage of the users browser and a copy is also stored on the server in the event that the local storage copy gets deleted.


     \noindent  If the file is available on the local storage, no request is sent to the server for the file. Even though the file is stored on a secure server, its safety is further guaranteed by the fact that it is stored in an encrypted format and the password to decrypt is possessed by no one other than the user. We have used AES block cipher encryption \cite{c10} to encrypt the file with the password derived from the fingerprint of the user.\smallskip

When the user wishes to perform operations like view records, create records or grant/revoke access to records, the user provides his/her fingerprint which is then used to decrypt the retrieved user public and private key. The private key can be used to decrypt only those records which were encrypted using the user’s public key. Correspondingly the public key can be used to encrypt fresh records. The public key of a user is accessible to all the other users of the system.

\subsubsection{Hospital SignUp}
Considering that a hospital cannot be registered using a fingerprint, a different approach would have to be followed for Hospital SignUp. The authenticity of a hospital would have to be tested physically by a committee following which a public-private key pair would be generated and assigned to the hospital. 
The hospitals will also be allowed to create a login password which in turn would be used to encrypt and decrypt the file containing the public key of the hospital stored in conjunction with the private key of the hospital.
As a result, there will exist a list of all the hospitals in the blockchain. This list will later be used for verification of transactions by smart contracts.

\subsection{Hospital Functionality}
A member of the hospital staff can log in using the hospital ID and password(this essentially decrypts the file containing the hospital public and private key). After this the hospital can either view records or enter a new record.
To submit a record the doctor fills out the record details. On submitting, the Patient ID is used to obtain \textit{P\textsubscript{Pub}} and this public key is used to encrypt the data. The encrypted record is then stored in IPFS while the SHA-256 hash retrieved from IPFS is stored in the smart contract storage along with the patient public key. Along with the hash, a signed copy of the hash(signed with the Patients Private Key) is sent to the smart contract as well. The contract first verifies the validity of the signature using the patients public key and also check whether the hospital public key is listed in the registered hospitals list. If both these conditions are validated, only then is the data stored in the smart contract. Only the patient can decrypt this record using  \textit{P\textsubscript{Pri}}. Before the record is stored in IPFS, the third smart contract is accessed and in the smart contract the count of the number of cases of the reported disease is increased by one. 

\textbf{\textit{IPFS Patient Record} $\gets$ \textit{Enc\textsubscript{\textit{P\textsubscript{Pub}}}}$\langle Newly Created Record \rangle$}

The hospital can view the records they have access to by entering the decryption password which decrypts the public private key combination file stored in the local storage of the browser. The unavailability of the record causes the system to fetch the same from the database. The \textit{H\textsubscript{Pri}} so obtained is used to decrypt the record. The hospital is required to enter the patient id of the patient whose records the hospital wishes to fetch. The \textit{id} is used to fetch the patient public key and the combined hospital patient public key is in turn sent to the smart contract which stores the mapping between the key and the record hashes. Once the desired entry in the map is found, the hash(es) are returned to the hospital which in turn uses the hash to fetch the record from the IPFS network.

If the patient revokes access, the hospital will no longer be allowed to view the record as the entry containing the IPFS hash will be removed from the smart contract.

\textbf{\textit{Hospital Record} $\gets$ \textit{Dec\textsubscript{\textit{H\textsubscript{Pri}}}}$\langle IPFS Hospital Record \rangle$}

\subsection{Patient Functionality}
A patient can log in to the application using their patient ID and password(fingerprint). Now the patient can view their records and can grant and revoke access to hospitals and other users. 

To view their records, the patient has to enter the decrypting password which is used to decrypt the public private key combination file stored in the local storage of the browser. The unavailability of the file on the browser causes the system to fetch the same from the database. \textit{P\textsubscript{Pri}} can then be used to decrypt the data as the encryption was done using the corresponding public key. \smallskip

\textbf{\textit{Patient Record} $\gets$ \textit{Dec\textsubscript{\textit{P\textsubscript{Pri}}}}$\langle IPFS Patient Record \rangle$} \smallskip

To grant access the user specifies the hospital ID and the list of records that the hospital should be given access to. These records are then decrypted and then re-encrypted using the \textit{H\textsubscript{Pub}} and stored in IPFS. The combined key, which is a concatenation of \textit{P\textsubscript{Pub}} and \textit{H\textsubscript{Pub}} is stored in the smart contract along with the IPFS hash. A signed copy of the hash(signed with the Patients Private Key) is sent to the smart contract as well. The contract first verifies the validity of the signature using the patients public key and if the signature is valid, only then is the data stored in the smart contract.

\textbf{\textit{IPFS Hospital Record} $\gets$ \textit{Enc\textsubscript{\textit{H\textsubscript{Pub}}}}$\langle Patient Record \rangle$} \smallskip

To revoke access, this entry is deleted from the smart contract. The patient sends the hospital patient combined key to the smart contract in order to delete the entry. The patient also has to send a signed copy of the combined key as well as the hospital key individually to the smart contract. The contract will first combine the hospital key with the patients public key and then verify the signature. If it is valid, the entry will be deleted. This prevents any random patient from deleting some other patient\textquotesingle s entry in the smart contract. The record is also deleted by the patient from his own system. This however does not ensure that the record is deleted all across the IPFS network because the hospitals that accessed the records have the data stored in their local systems(because of how IPFS works). Thus, the patients must also send a notification to the concerned hospitals to 
delete the corresponding copies of the particular record from their local systems as well. \textit{The patients can then check if anyone still has copies of the file by using the command :- ipfs dht findprovs $\langle file hash \rangle$ }. If any of the hospitals whose access has been revoked pop up, legal action can be taken against them.

\subsection{Statistical Analysis}
It is very important to maintain the statistics and making the information available to the public without violating the privacy of the patients. The proposed model has the provision to maintain the statistics without violating the patient privacy. Hospitals posting health record hash to the smart contract should also update the disease count.\smallskip

Consider two records \textit{R\textsubscript{0}} and \textit{R\textsubscript{1}}. Let the diseases updated by the records be \textit{D\textsubscript{0}} and \textit{D\textsubscript{1}} respectively. Let there be an attacker \textit{A} such that \textbf{\textit{A$\langle R\textsubscript{i} \rangle$} = \textit{D\textsubscript{i}}} represent that the attacker correctly establishes the disease count transaction that corresponds to a particular record storage transaction. In order to ensure that the attacker cannot conclusively establish the mapping \textbf{Pr$\langle \textit{A$\langle R\textsubscript{i} \rangle$ = D\textsubscript{i}} \rangle$ = 1/2}. To ensure this, the hospital must update the count of atleast two diseases at a time. However there are the following issues in maintaining the privacy\smallskip

\textbf{\textit{Gender based disease distinction}:} If a hospital updates for two patients (female and male) say pregnancy and diabetes respectively. It can be easily mapped that the female is pregnant and male is having diabetes. \smallskip

\textit{Solution}: Since the blockchain does not hold the patient identity details, establishing a relation between the patient and disease will not be possible.\smallskip

\textbf{\textit{Specialized Hospital}:} If any hospital is specialized for treatment of a particular disease then it can be more easily identified because it always updates the count for the same disease. \smallskip

\textit{Solution}: Even though patient blockchain address and personal details cannot be mapped, the blockchain address and disease can be mapped. However, there are cases where patient may not have the disease. For example, if we say hospital posts 10 hash transaction and 9 disease count update to blockchain then disease and blockchain address cannot be mapped. It gives high probability but it is non-deterministic.

To access the statistical information, the third smart contract is being used in the system. It can be used to retrieve the number of cases of each disease that have been encountered. It basically stores a mapping between the name of the disease and the number of records reporting that disease. It does not contain any details about the patients having the disease whatsoever. The functionality can be extended to allow number of cases of a particular disease to be reported location wise as well. In that case, the map would contain the location and the disease being mapped to the number of cases of the particular disease.

\section{ALGORITHMS USED}

Algorithm 1 is used for storing the IPFS hash in the smart contract. Once the hash for an IPFS record is obtained, it can not be stored in the smart contract directly because it is too large to be converted to the Bytes32 format. Thus, we first split the hash into two parts. After this, a random character is added at the end of each part as a security measure. Following this, the two parts of the hash are converted into the Bytes32 format and then correspondingly stored in the smart contract.

\begin{algorithm}[H]
\caption{Storing the IPFS hash in the smart contract}\label{euclid}
 \hspace*{\algorithmicindent} \textbf{Input:} IPFS Hash of the record 
\begin{algorithmic}[1]
\State $\textbf{\textit{ihash}} \gets \text{IPFS hash of the record}$
\State $\textbf{\textit{p1}} \gets \text{ihash.substring$\langle0,24\rangle$}$
\State $\textbf{\textit{p2}} \gets \text{ihash.substring$\langle25,48\rangle$}$
\State $\textbf{\textit{p1\_r1}} \gets \text{\textit{p1}+randomCharacter}$
\State $\textbf{\textit{p2\_r2}} \gets \text{\textit{p2}+randomCharacter}$
\State $\textbf{\textit{finalpart1}} \gets \text{convertToBytes32(\textit{p1\_r1})}$
\State $\textbf{\textit{finalpart2}} \gets \text{convertToBytes32(\textit{p2\_r2})}$
\State Store \textbf{\textit{finalpart1}} and \textbf{\textit{finalpart2}} in smart contract with a mapping to the combined key/patient key
\end{algorithmic}
\end{algorithm}

Algorithm 2 explains the process followed to retrieve the hashes from the smart contract and correspondingly convert them to a usable IPFS hash. The data stored in the smart contract is in Bytes32 format. The smart contract returns the hash of a record as a single string with the two parts of the hash separated by a comma. We first split the string on the comma. Then each of the parts obtained is first converted into the Base58 format. Following this, the random character that was added at the end of each part while storing it in the smart contract is removed. The two parts are then concatenated to obtain the original IPFS hash of the record.

\begin{algorithm}[H]
\caption{Retrieving the IPFS hash from the smart contract}\label{euclid}
 \hspace*{\algorithmicindent} \textbf{Input:} Patient public key/ Combined key\\
 \hspace*{\algorithmicindent} \textbf{Output:} IPFS Hash of the record
\begin{algorithmic}[1]
\State $\textbf{\textit{contracthash}} \gets \text{hash retrieved from the smart contract}$
\State $\textbf{\textit{finalpart1}} \gets \text{contracthash.splitonComma[1]}$
\State $\textbf{\textit{finalpart2}} \gets \text{contracthash.splitonComma[2 ]}$
\State $\textbf{\textit{p1\_r1}} \gets \text{convertToBase58(finalpart1)}$
\State $\textbf{\textit{p2\_r2}} \gets \text{convertToBase58(finalpart2)}$
\State $\textbf{\textit{p1}} \gets \text{\textit{p1\_r1}-r1}$
\State $\textbf{\textit{p2}} \gets \text{\textit{p2\_r2}-r2}$
\State $\textbf{\textit{hash}} \gets \text{\textit{p1}+\textit{p2}}$
\end{algorithmic}
\end{algorithm}

\section{OPTIMISATION OF STORAGE ON BLOCKCHAIN}
In the method proposed above, we have used smart contracts to store our mapping between patient/combined key and the ipfs hash values. This works for us because of the fact that our system uses a \textit{Permissioned Blockchain} instead of a Public Blockchain as in the case of the latter, the cost to store such a large amount of data in the smart contract would render it infeasible. Smart contracts were not designed for the purpose of storing large amounts of data as storing large amounts of data on the blockchain would essentially slow down and add additional load on the entire system. Hence, we have suggested an alternative mechanism where we use IPFS to store the patient public key to hash mapping instead of the smart contract

In this method, the patient public key to hash mapping is stored in the IPFS network. The hash of this map is stored in a smart contract. Whenever a hospital or patient needs to retrieve the hash of a particular record, they first fetch the hash of the map from the smart contract. After this, they can access the record hash using the patient/combined key. The primary advantage of this method is that the only thing the smart contract stores is the hash of the map.\smallskip

There will be a total of three smart contracts in this method.
\begin{enumerate}
\item A smart contract storing the hash of the map stored in the IPFS network. This smart contract is also used by a patient to view his/her records.
\item A smart contract which is accessed by the patient when they want to grant or revoke access to a hospital.This smart contract also stores the combined patient and hospital key to hash mapping.
Revoking access would simply delete the required entry in the map while granting access would store the hash of the newly encrypted record in the mapping.
\item A smart contract which is accessed when a hospital creates a particular record and need to store the corresponding record in IPFS. This smart contract in turn accesses the first smart contract to get the hash of the map.It then accesses the map and adds the required entry to the map.The hash of the updated mapping is then stored back in the first smart contract.
\end{enumerate}

The reason we transfer only one of the mappings to IPFS is that the other two maps in our system i.e. the combined key to hash map and the disease name to occurences of disease map both involve entries that need deletion and updation. While we can add additional things to a record in IPFS and store it efficiently using data deduplication, that will not be possible when deletions and updations are required. Thus every change to the map will result in the creation of a new map in the system which will lead to the wastage of a lot of space.
\section{RESULTS}
The record creating/viewing interface has been created using React js and the backend is pure javascript. Mongo DB has also been used to store the patient and hospital ids and their public keys.
In order to create a simulation of  blockchain network, we have used ganache-cli as our test blockchain. We have also used Web3 js to interact with the blockchain.
For testing the IPFS network we have used INFURA, which provides a  secure, reliable and scalable access to the IPFS gateway. It provides TLS enabled IPFS gateways which can in turn be used to access up and running IPFS nodes. \smallskip

The test model has been implemented on a system having 2.2 GHz Intel Core i5-5200U processor with Intel HD Graphics 5500,
3 MB cache, 2 cores and a 1 TB 5400 rpm SATA Hard Drive.
For generating the results, the MongoDB database was hosted on AWS(Amazon Web Services) using mlab. This gave us an idea of what a standard cloud service provider response time would be like. The network being used was a standard 4G connection.\smallskip



\begin{figure}[H]
\includegraphics[width=0.5\textwidth,height = 6cm]{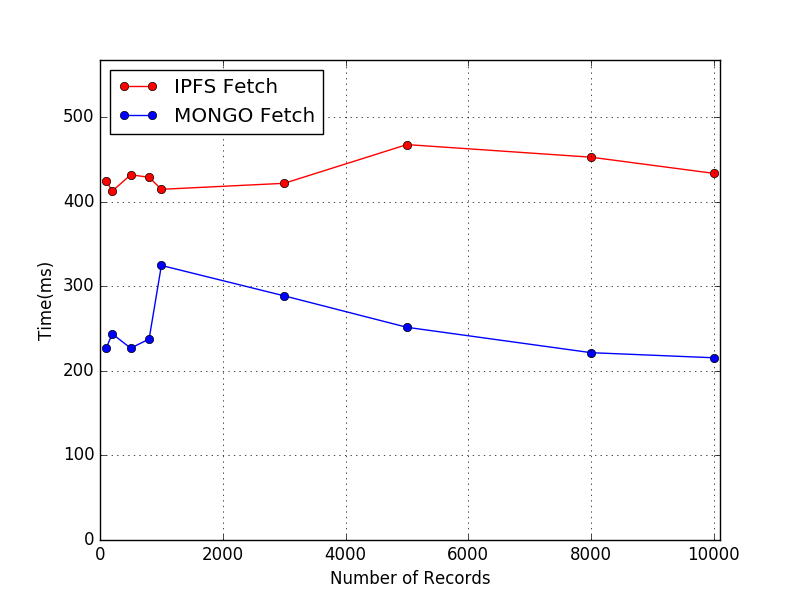}
\caption{\label{fig:The-caption}Record Retrieval Time Comparison}
\end{figure}

The results in Fig 3 give a time vs number of records comparison when using IPFS and also if MongoDB was used in place of IPFS to store records. The fetch speed from IPFS depends on a number of factors such as the number of systems storing the records, location of the closest system containing the record etc. Thus, it is not possible to properly analyse the IPFS fetch time on a simulation of the actual system such as ours.

Even if the fetch speed is slightly slower than a traditional cloud service, IPFS is still the ideal choice for our model because it provides immutability of records and decentralisation. Fig. 4 shows the amount of time that it takes to retrieve hashes of records from the smart contracts while varying the number of participants in the network. Considering that this has been tested using ganache-cli, the variation in retrieval time with the number of participants is not substantial. 

\begin{figure}[H]
\includegraphics[width=0.5\textwidth,height = 6cm]{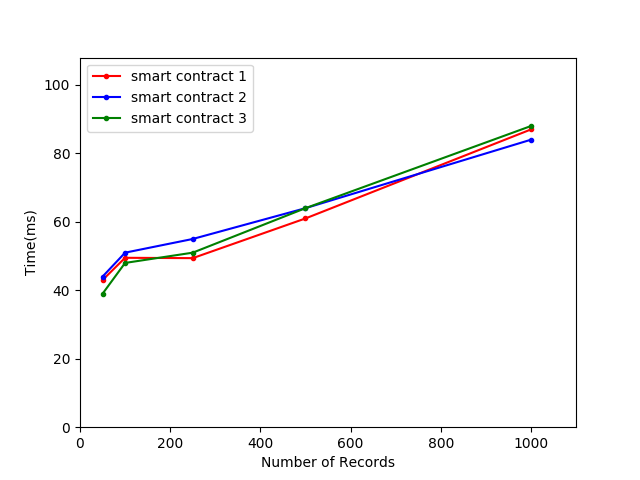}
\caption{\label{fig:The-caption}Hash Retrieval from Smart Contract Time Comparison}
\end{figure}

\section{DISCUSSION and LIMITATIONS}
While the main e-Health requirement is confidentiality, this is also one of the main challenges. Although the proposed system prevents the hospital from gaining access to the record without the patient's permission, there is no mechanism to prevent the hospital from taking a photograph or copying the contents of the record when uploading the record or during the time they have access to the record. 

Another challenge is that in a developing countries it is difficult to enforce a digital system as many people do not have access to the internet or would be unaware of how to use such a system \cite{c8}. Additionally, in IPFS, if data is to be shared to other systems, the system containing the record must be online failing which the transfer will not be possible. If multiple systems contain the record, at least one of them must be online for record transfer to be possible. 

The primary advantage of using blockchain in this scenario is that it provides a completely decentralized and immutable logging system in which everyone can view who gave permission to a particular hospital, who revoked access of a particular hospital, who created or accessed which record etc. The fact that this immutable logging  of data is available to all the participants in the blockchain makes it more secure and misuse of data and records can be easily identified. The use of IPFS along with blockchain makes it a  completely decentralised system which does not require on any organisation for logging or storing of data. 
IPFS ensures immutability of patient records while blockchain ensures the immutability of logged transactions. The use of biometric encryption ensures that even in a scenario in which a patient is in a critical condition and can not provide access to his/her records, the records could still be accessed using his/her fingerprints.
However, the following points need to be taken into consideration which might not make Blockchain and IPFS(in it's current form) the best technologies for such a system :-
\begin{itemize}
    \item In IPFS, the data is initially stored on the system that adds it to the IPFS network. When the data is accessed by other systems, the data is sent through the network to those systems as well which in turn can the act as the provider of the data. The more widespread the data is, the faster is the performance of the system. In this scenario however, the data is originally stored with the hospital that creates it and the only person who will ever fetch the data is the patient. Thus, in a normal scenario, every record will have a replication factor of 2 at maximum which will not allow the IPFS network to function to its potential. To avoid this, a plausible solution would be to configure IPFS to have a replication factor of atleast 20 for each record.
    \item The fact that the Blockchain will keep a track of which patient visited which hospital can be considered as a privacy infringement in itself as each user's hospital visit is being broadcasted over the network. It should however be considered that all the users are identified using their public keys in the system and their names are not exposed to the people on the system. Thus, correlating which public key corresponds to which patient will not be an easy task.
    \item It is also impossible for a user to delete all copies of a record on an IPFS network by himself/herself. Even though the user can track who stores the copy of the records even when the hash has been deleted from the smart contract, this does not seem like an ideal solution to the problem. 
    \item There is also the fact that a transaction containing the storage of the deleted hash in the smart contract will also always exist as a part of the blockchain. Thus, with a considerable amount of effort, a hospital might be able to regain the hash of the record that the patient deleted(when revoking the access of the hospital). A viable solution for this particular problem could be using \textbf{IPNS(Inter Planetary Naming System)}\cite{c16} records to point to such records for which a user might have to revoke access to later. As a result, only the hash of the IPNS record will be registered in the blockchain and not the hash of the medical record directly.
    \item Storage of data on a Blockchain is also not recommended. It would greatly limit the extent to which such a mechanism can scale. Considering that IPFS cannot be used to store the contents of both the smart contracts, a working solution might involve using cloud storage which would damage the decentralised nature of the system.
    \item In the current version of our system, the hospital exists as a single entity on the blockchain. This, however, is not very appropriate as a hospital would have several departments with several doctors in each department and one particular department or doctor should not be able to create or view the records of another department without permission.
\end{itemize}
\section{FUTURE WORKS}
The above proposed system has currently been tested using ganache-cli and the IPFS test nodes provided by INFURA. However, in order to realise the full potential and limitations of the system it needs to be tested on an actual permissioned blockchain with a proper IPFS network setup. A proper setup will further help us evaluate how the system behaves when millions of users try to store IPFS data, access and update smart contracts, mappings stored on the IPFS network, etc simultaneously.

\section{Acknowledgements}
We would like to thank Indian Institute of Information Technology, Allahabad for extending support to the first author through a seed money project titled “Electronic Medical Record Management System using Blockchain Technology”.

\addtolength{\textheight}{-9.5cm}

\end{document}